\documentclass{article}
\usepackage{PRIMEarxiv}
\usepackage{amsmath, amssymb}
\usepackage{graphicx}
\usepackage{xcolor}
\usepackage{graphicx}
\usepackage{amsmath,amsfonts,amssymb,mathrsfs}
\usepackage{caption}
\usepackage{booktabs}
\usepackage{subcaption}
\usepackage{hyperref}
\usepackage[utf8]{inputenc}
\usepackage{algorithm}
\usepackage[noend]{algpseudocode}
\usepackage{tabularx} 
\usepackage{cite}

\newcommand{\lefteqone}[1]{%
  \makebox[0.9\linewidth][l]{\hspace{6em}#1}%
}
\newcommand{\lefteqtwo}[1]{%
  \makebox[0.9\linewidth][l]{\hspace{3em}#1}%
}

\newcommand{\bbGamma}{{\mathpalette\makebbGamma\relax}}
\newcommand{\makebbGamma}[2]{%
  \raisebox{\depth}{\scalebox{1}[-1]{$\mathsurround=0pt#1\mathbb{L}$}}%
}

\title{High-fidelity level-set modeling of polycrystalline \; grain growth}
\author{
  Tianchi Li, Marc Bernacki \\
  Mines Paris, PSL University\\
  Centre for material forming (CEMEF), UMR CNRS\\
  06904 Sophia Antipolis, France\\
  \texttt{tianchi.li@minesparis.psl.eu, marc.bernacki@minesparis.psl.eu} \\
}

\begin{document}
\maketitle

\begin{abstract}
Accurate modeling of polycrystalline microstructure evolution under strong crystallographic heterogeneities remains a major challenge for full-field numerical methods at the mesoscopic scale. In this work, we present a high-fidelity level-set framework for capillarity-driven grain growth in polycrystals with highly-heterogeneous, disorientation-dependent grain boundary energies. The novel framework represents a polycrystalline extension of our level-set formulation, previously developed and validated using a single triple junction benchmark case. In-depth comparisons with three established level-set models demonstrate that the proposed method yields the most energetically-consistent evolution of grain statistics, disorientation distribution function, and triple junction dihedral angles. Accuracy and robustness are maintained across the entire heterogeneity spectrum. To the best of our knowledge, this approach delivers the highest-fidelity front-capturing level-set modeling of grain growth based on Mullins' mean curvature flow theory, paving the way for state-of-the-art digital twins for annealing applications.
\end{abstract}


$$
$$

Annealing has long been practiced in blacksmithing to restore ductility in deformed metals. However, it was not until the early 20\textsuperscript{th} century that the grain coarsening occurring after recrystallization, now referred to as grain growth, was identified as a distinct metallurgical phenomenon through systematic metallographic observations \cite{humphreys_2017}. By the mid-20\textsuperscript{th} century, capillarity (surface tension) was recognized as the primary driving pressure for grain growth \cite{smith_1948_rohrer}, enabling the development of quantitative kinetic models \cite{burke_1952,von_neumann_1952}. In 1956, Mullins formulated the classical mean curvature flow equation for capillarity-driven grain boundary (GB) migration \cite{mullins_1956}:
\begin{equation} \label{Mullins}
    \vec{v} = - \mu \gamma \kappa \vec{n} ,
\end{equation}
where $\vec{v}$, $\mu$, $\gamma$, $\kappa$, $\vec{n}$ denote respectively the velocity, mobility, free energy, mean curvature, and unit outward normal vector of GBs. Despite its simplicity, Eq. \eqref{Mullins} successfully predicts the evolution of mean grain size and even grain size distribution during grain growth \cite{hillert_1965}. This represents a core interest for grain growth modeling since the eventual grain size strongly influences the mechanical and functional properties of metallic components \cite{lasalmonie_1986,zhang_2022}. However, the validity of Eq.~\eqref{Mullins} was questioned in recent microscopic studies focusing on individual GBs \cite{bhattacharya_2021,florez_2022,muralikrishnan_2023,xu_2024}. As discussed by Yang \cite{yang_2026} and Bernacki \cite{bernacki_2025}, these discrepancies can arise from the neglect of GB heterogeneity and anisotropy in the current framework. Nevertheless, for large polycrystalline systems, Mullins’ equation remains statistically predictive of average grain growth behaviors, as proven by its longstanding success in annealing predictions. 

From a thermodynamic perspective, grain growth is driven by the reduction of total interfacial energy through the decrease of GB area. As derived by Herring \cite{herring_1999}, the first order capillarity driving pressure for an infinitesimal surface displacement is given by
\begin{equation} \label{Herring}
P \approx - \underbrace{\left( \gamma \mathbb{I} + \nabla_{\vec{n}} \nabla_{\vec{n}} \gamma \right)}_{\bbGamma(\vec{n})} : \mathbb{K},
\end{equation}
with $\mathbb{K}=\nabla\vec{n}$ the curvature tensor and $\bbGamma$ the GB stiffness tensor. When $\gamma$ is assumed independent of interface inclination ($\nabla_{\vec{n}} \gamma = 0$), the driving pressure simplifies to the one used in Eq. \eqref{Mullins}: $P \approx - \gamma \kappa$ \cite{bernacki_2024}. This simplification excludes the full anisotropic GB energy description \cite{bulatov_2014}, thereby limiting the applicability of the traditional framework. To address these limitations, advanced level-set formulations have been proposed \cite{murgas_2021}. These approaches can be broadly categorized as 3-parameter heterogeneous models \cite{fausty_2018,fausty_2020}, in which $\gamma$ depends only on disorientation, and 5-parameter anisotropic models \cite{hallberg_2019,fausty_2021}, where $\gamma$ depends on both disorientation and inclination. By including the spatial derivative of $\gamma$, the new models aim to better capture the influence of GB heterogeneity on triple junction (TJ) kinetics compared to the direct substitution of disorientation-dependent GB energy in Mullins' equation \cite{jin_2015}. Despite increased complexity and computational costs associated to additional gradient terms, their overall performance remains limited, particularly in polycrystals with highly heterogeneous GB energies. 

In a previous study \cite{li_2025}, we introduced a novel level-set formulation for capillarity-driven GB migration. The model accurately reproduces the velocity, dihedral angles, and equilibrium profile of a single TJ over a wide range of GB energy ratios, including extreme cases of wetting and complete stagnation. Although restricted to disorientation-dependent interface energies, the framework demonstrates strong potential for capturing realistic GB kinetics. In the present work, we extend this formulation to the polycrystalline scale and perform large-scale grain growth simulations under diverse GB energy configurations. The performance of the approach is systematically assessed through comparison with existing level-set models, focusing on the evolution of global statistics, disorientation angle distribution, and TJ dihedral angles. 

\begin{figure}[ht]
    \centering
    \includegraphics[width=0.7\textwidth]{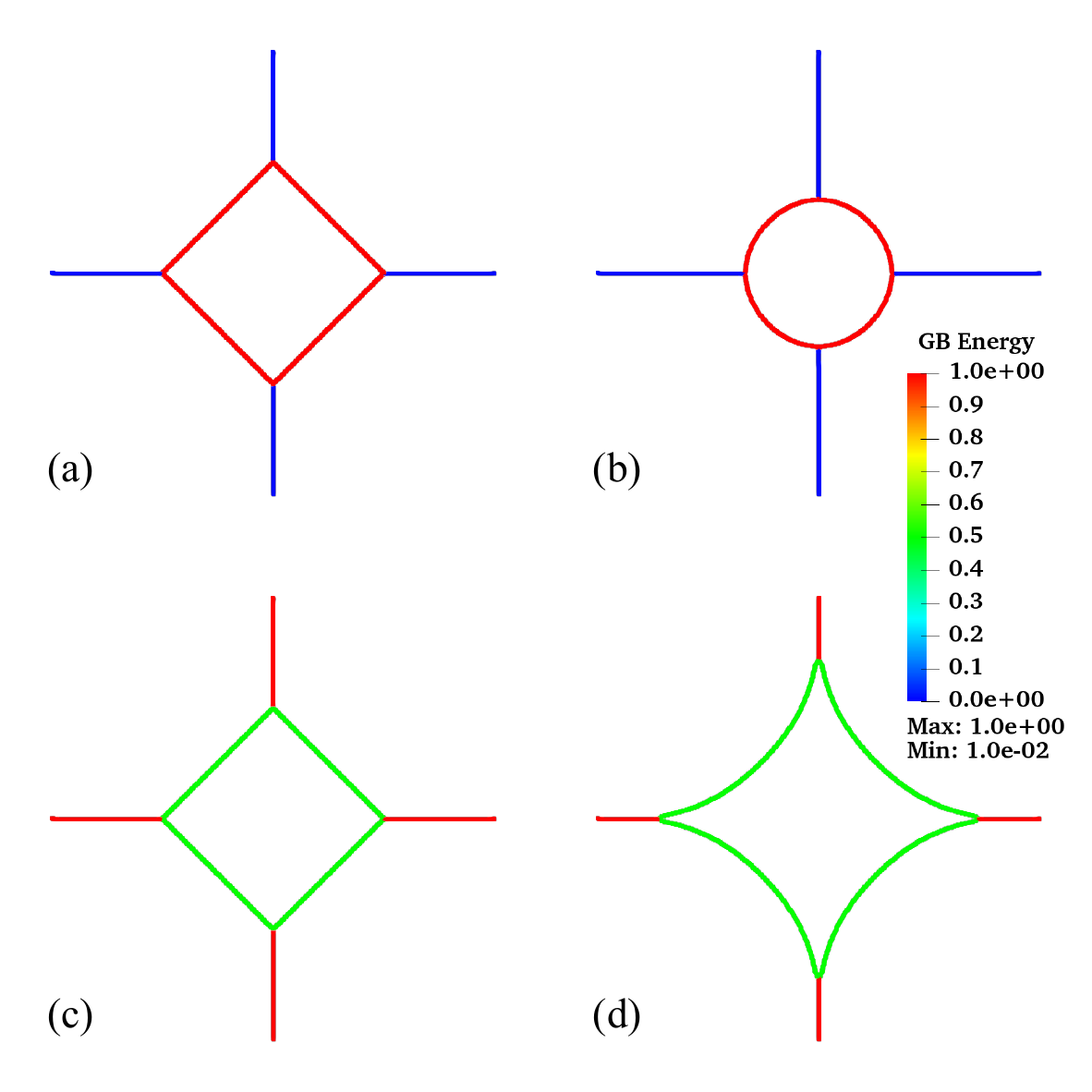}
    \caption{Capillarity-driven grain growth strongly influenced by grain boundary (GB) heterogeneity. $R_\gamma$ denotes the free-energy ratio between the 4 inner and 4 outer GBs. \textbf{(a)} Initial configuration and \textbf{(b)} evolved grain morphology for $R_\gamma = 100$. \textbf{(c)} Initial configuration and \textbf{(d)} evolved grain morphology for $R_\gamma = 0.5$.}
    \label{Figure_1}
\end{figure}

The proposed formulation fundamentally differs from all other established level-set frameworks in the introduction of crystallographic heterogeneities. Rather than including a convective term based on the spatial gradient of GB energy $\gamma$, one accounts for the disorientation dependency of $\gamma$ through a source term on the right-hand side of the level-set transport equation:
\begin{equation} \label{LS_transport_Tianchi}
    \frac{\partial \psi_i}{\partial t} - M_R\Delta\psi_i = \Lambda_i \left( 1 - \sum_{j=0}^{N-1} H(\psi_j) \right) .
\end{equation}
Here, $\psi_i$ denotes the $i$-th global level-set distance function obtained via coloring techniques \cite{scholtes_2015}. $\Lambda_i$ is the source term factor field associated with $\psi_i$, and $H$ is a slightly regularized Heaviside function \cite{zhao_1996}. The diffusive term on the left-hand side preserves the classical curvature-driven mechanism, governed by the reduced mobility $M_R=\mu\gamma$. To facilitate numerical implementation, the source term is expressed as $\Lambda_i=\lambda_i\max{(\Lambda_i)}$, with $\lambda_i$ the normalized source term factor and $\max{(\Lambda_i)}$ a magnitude constant. For finite element simulations using unstructured triangular mesh, the optimal parameter relations are obtained through dimensional analysis: 
\begin{equation} \label{optimal_parameters}
    \max{(\Lambda_i)} \simeq \sqrt{\frac{M_R}{\Delta t}} , \qquad h \simeq \sqrt{M_R \Delta t} ,
\end{equation} 
with $\Delta t$ the simulation time step and $h$ the minimum refined mesh size at TJs. Once the average reduced mobility $M_R$ is prescribed, the remaining parameters follow from the chosen spatial or temporal resolution. $\lambda_i$ can be interpolated from the GB energy configuration at TJs using the dataset reported in \cite{li_2025}. In contrast to other existing level-set formulations, the present approach does not involve computationally expensive operations. Nevertheless, it accurately captures the influence of polycrystalline heterogeneity on the migration of individual interfaces and junctions during capillarity-driven grain growth, as illustrated in Fig. \ref{Figure_1}. \\

\begin{figure}[ht]
    \centering
    \includegraphics[width=\textwidth]{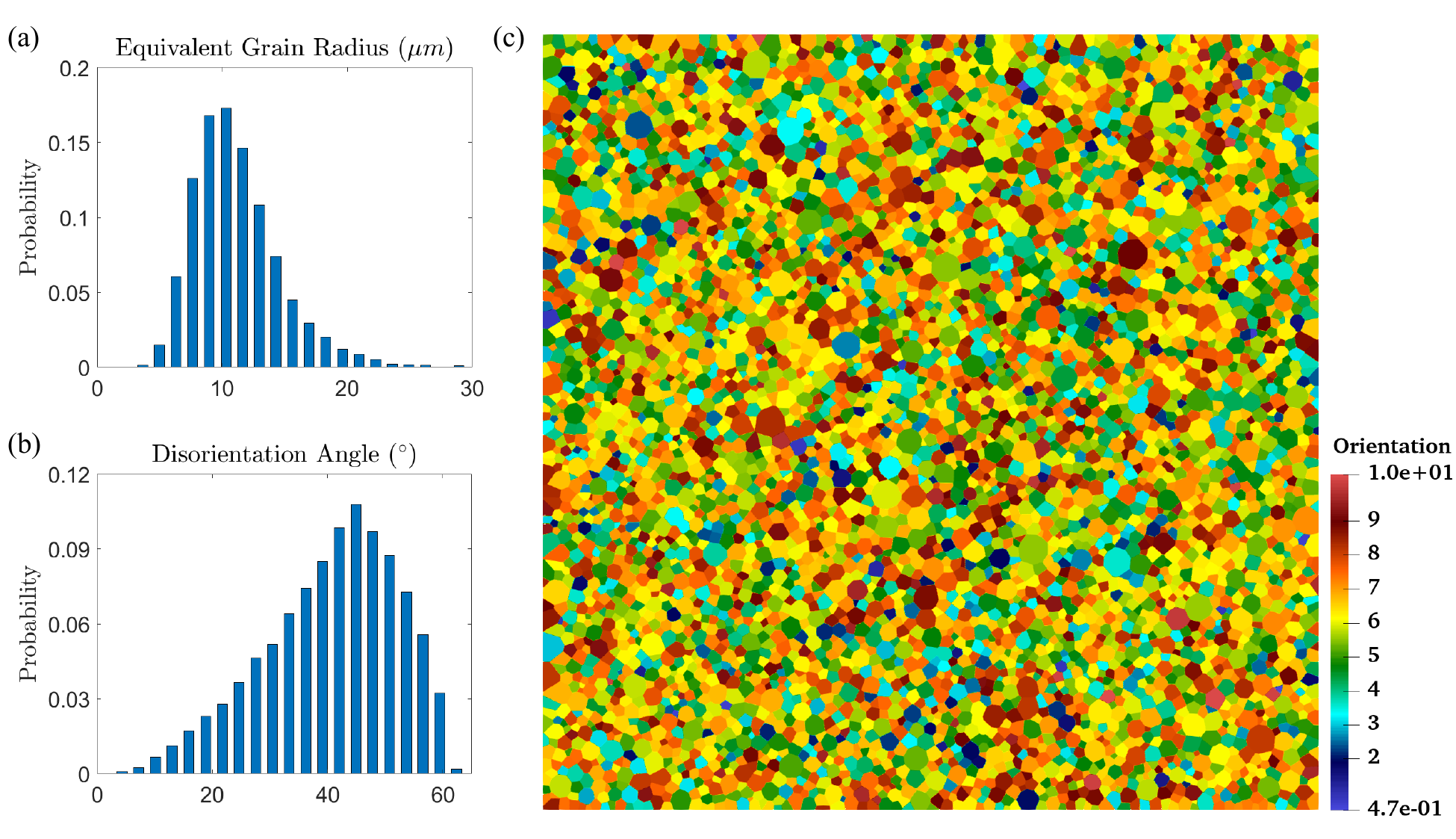}
    \caption{Polycrystalline system used for simulations in this work. \textbf{(a)} Grain size distribution. \textbf{(b)} Disorientation angle distribution. \textbf{(c)} Orientation magnitude map.}
    \label{Figure_2}
\end{figure}

To assess the performance of the proposed formulation at the polycrystalline scale, an initial microstructure with a log-normal grain size distribution (Fig.~\ref{Figure_2} (a)) is generated in a $1.5 \times 1.5 \ \text{mm}^2$ square domain using Laguerre–Voronoi tessellation \cite{hitti_2012}. Crystallographic orientations are assigned by randomly generating the Euler angles $(\varphi_1,\Phi,\varphi_2)$ for all grains, resulting in a Mackenzie-type disorientation angle distribution \cite{mackenzie_1958} (Fig. \ref{Figure_2} (b)). A graphical representation of the polycrystal is shown in Fig. \ref{Figure_2} (c) using a color scheme based on orientation magnitude $\sqrt{\varphi_1^2+\Phi^2+\varphi_2^2}$. 

\begin{figure}[ht]
    \centering
    \includegraphics[width=0.73\textwidth]{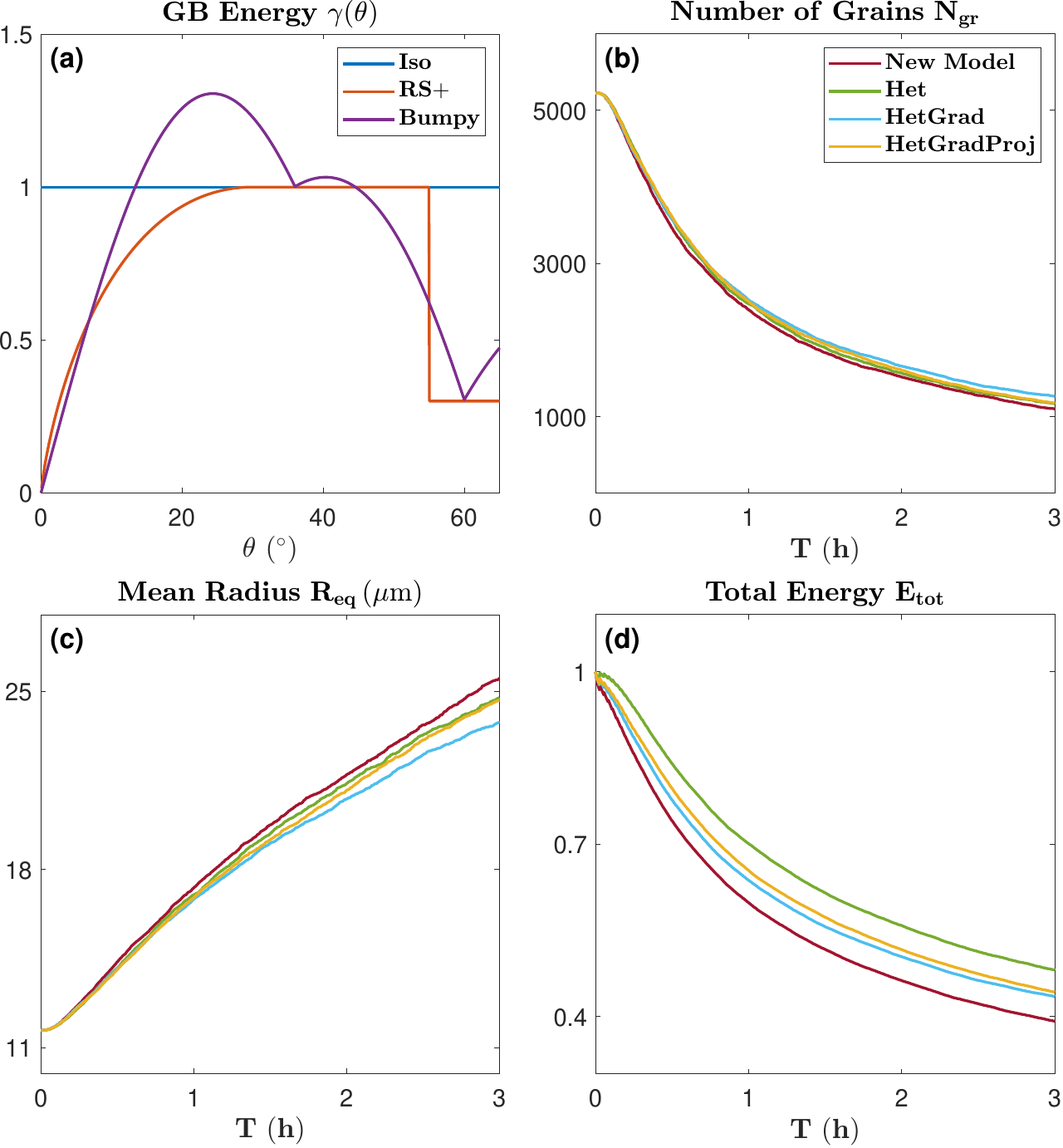}
    \caption{GB energy functions considered in the current work and grain growth statistics measured in the Bumpy case. \textbf{(a)} The analytical forms of $\gamma(\theta)$ and the evolution of \textbf{(b)} number of grains, \textbf{(c)} mean equivalent grain radius, and \textbf{(d)} total normalized GB energy in Bumpy simulations, with comparison to existing level-set models.}
    \label{Figure_3}
\end{figure}

As in \cite{li_2025}, the GB energy $\gamma$ is treated as dimensionless, since it enters the formulation only through the normalized source-term factor. Assuming that $\gamma$ depends solely on the disorientation angle $\theta$, three analytical GB energy functions are considered (Eqs.~\eqref{Iso}–\eqref{Bumpy}, Fig.~\ref{Figure_3} (a)). The Iso case corresponds to uniform GB energy and serves as a reference configuration. RS$^+$ extends the prevalent Read-Shockley model \cite{read_shockley_1950} by introducing a low-energy regime at high disorientation angles to mimic twin boundaries. However, both isotropic and Read-Shockley-type descriptions yield nearly constant energies in the mid-to-high disorientation range (30$^\circ$ to 55$^\circ$), which dominates the Mackenzie distribution. To probe strongly heterogeneous scenarios, a Bumpy function with amplified bumps and cusps is introduced, providing a stringent test of numerical robustness. \\

\textbf{Iso:}
\begin{equation} \label{Iso}
\lefteqone{%
$\begin{aligned}
\gamma &= 1 \, ,
\end{aligned}$} 
\end{equation}

\vspace{1em}

\textbf{RS$^+$:}
\begin{equation} \label{RS+}
\lefteqone{%
$\begin{aligned}
\gamma &=
\begin{cases}
\dfrac{\theta}{\theta_{\max}}
\left(1 - \ln\!\left(\dfrac{\theta}{\theta_{\max}}\right)\right),
& \theta < \theta_{\max} \\
1, & \theta_{\max} < \theta < \theta_{\mathrm{thresh}} \\
0.3, & \theta > \theta_{\mathrm{thresh}} 
\end{cases} \, ,
\end{aligned}$} 
\end{equation}

\noindent\hspace{4em}%
where $\theta_{\max}=30^\circ$ and $\theta_{\mathrm{thresh}}=55^\circ$,

\vspace{2em}

\textbf{Bumpy:}
\begin{equation} \label{Bumpy}
\lefteqone{%
$\begin{aligned}
\gamma &=
\gamma_b \left( \alpha |\sin(3\theta)| + \beta |\sin(5\theta)| \right) \, ,
\end{aligned}$} 
\end{equation}

\noindent\hspace{4em}%
where $\gamma_b=1.17$, $\alpha=0.9$, $\beta=0.3$. \\

\vspace{1em}

Three existing level-set formulations are selected for comparison (Eqs. ~\eqref{transport_het}–\eqref{transport_hetgradproj}). The Heterogeneous (Het) model directly incorporates $\gamma(\theta)$ into Mullins’ equation \cite{jin_2015}. Although $\gamma$ is constant along each individual GB, its discontinuity across junctions induces local heterogeneity. The Heterogeneous with Gradient (HetGrad) formulation captures this heterogeneity effect using the spatial gradient of $\gamma$ \cite{fausty_2018}. Since $\gamma$ is only defined on the GBs, one can further project $\vec{\nabla}\gamma$ along the interfaces to obtain the so-called Heterogeneous with Projected Gradient (HetGradProj) formulation \cite{murgas_2021}. 
\begin{equation} \label{transport_het}
\lefteqtwo{%
$\begin{aligned}
&\textbf{Het:} \qquad \qquad \quad \;\, \frac{\partial\psi}{\partial t} -\mu\gamma(\theta)\Delta\psi = 0
\end{aligned}$} \vspace{0.1em}
\end{equation}
\begin{equation} \label{transport_hetgrad}
\lefteqtwo{%
$\begin{aligned}
&\textbf{HetGrad:} \qquad \quad \, \frac{\partial \psi}{\partial t} - \mu\gamma(\theta)\Delta\psi + \mu\Vec{\nabla}\gamma(\theta) \cdot \Vec{\nabla}\psi = 0
\end{aligned}$} 
\end{equation}
\vspace{0.1em}
\begin{equation} \label{transport_hetgradproj}
\lefteqtwo{%
\hspace{0.8em}$\begin{aligned}
&\textbf{HetGradProj:} \quad \; \frac{\partial \psi}{\partial t} - \mu\gamma(\theta)\Delta\psi + \mu\Vec{\nabla}_{\Vec{n}}\gamma(\theta) \cdot \Vec{\nabla}\psi = 0
\end{aligned}$} 
\end{equation} 

In the present model, consistent with the dimensionless treatment of $\gamma$, the diffusive term is evaluated with $\gamma=1$. The reduced mobility $M_R$ in Eq.~\eqref{LS_transport_Tianchi} therefore coincides with the GB mobility $\mu$ in Eqs.~\eqref{transport_het}–\eqref{transport_hetgradproj}. Spatial discretization is performed using an unstructured triangular mesh with isotropic refinement near GBs. Considering the initial mean grain radius $\bar{R}=11$ \textmu m and a total simulation time of $T=3$ h, the minimum mesh size and time step are set to $h = 1$ \textmu m, $\Delta t = 10 \; \text{s}$. The remaining parameters follow from Eq.~\eqref{optimal_parameters}, yielding
\begin{equation}
     \quad M_R = \mu = 0.1 \; \text{\textmu m}^2 \, \text{s}^{-1}, \quad \max{(\Lambda_i)} = 0.1 \; \text{\textmu m} \, \text{s}^{-1} .
\end{equation}
Grain growth simulations are performed with the four level-set formulations described above, using an in-house finite-element library \cite{digonnet_2007,mesri_2009,bernacki_2011}. The temporal evolution of the number of grains $N_{\mathrm{gr}}$, the mean equivalent grain radius $R_{\mathrm{eq}}$, and the total normalized GB energy $E_{\mathrm{tot}}$ are recorded and reported in Fig.~\ref{Figure_3} (b–d). To maximize the influence of interfacial heterogeneity, the Bumpy energy function is selected for this comparison. Across all three metrics, the proposed formulation exhibits the fastest and most coherent evolution. In contrast, the other existing models display mutually inconsistent trends. The Het formulation shows the slowest reduction of $E_{\mathrm{tot}}$, while predicting faster changes in $N_{\mathrm{gr}}$ and $R_{\mathrm{eq}}$ than both HetGrad and HetGradProj. HetGrad yields the slowest evolution in $N_{\mathrm{gr}}$ and $R_{\mathrm{eq}}$, yet reduces total GB energy more rapidly than Het and HetGradProj. These discrepancies indicate a lack of energetic consistency in the existing formulations. Overall, the proposed framework provides the most energetically consistent grain growth kinetics among the four models.

\begin{figure}
    \centering
    \includegraphics[width=0.75\textwidth]{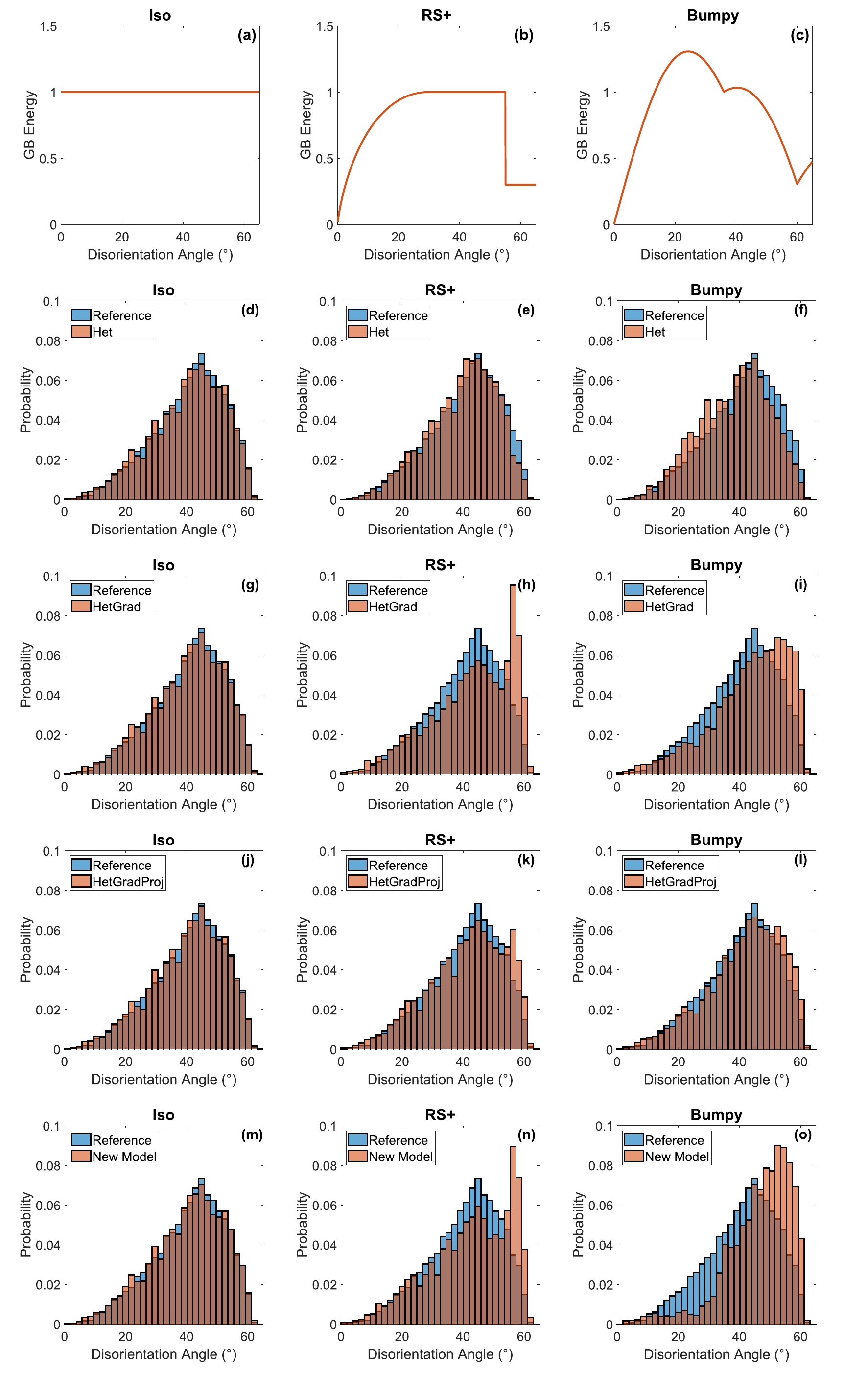}
    \caption{Disorientation Distribution Function (DDF) weighted by grain boundary (GB) length after 2 hours of grain growth simulation using \textbf{(a)} Iso, \textbf{(b)} RS$^+$, and \textbf{(c)} Bumpy GB energy functions, with comparison against initial DDF. Compared to \textbf{(d)}\textbf{(e)}\textbf{(f)} Het, \textbf{(g)}\textbf{(h)}\textbf{(i)} HetGrad, \textbf{(j)}\textbf{(k)}\textbf{(l)} HetGradProj formulations, \textbf{(m)}\textbf{(n)}\textbf{(o)} the new model gives the most energetically coherent DDF prediction.}
    \label{Figure_4}
\end{figure}

The global grain statistics can be further elucidated through the analysis of disorientation distribution function (DDF). To this end, grain growth simulations are performed using all four formulations and the three GB energy functions. Figure~\ref{Figure_4} compares the DDF measured at $T=2$ h with the initial reference distribution for each of the twelve cases. In the isotropic cases (Fig.~\ref{Figure_4}, left column), all four models preserve the Mackenzie distribution, as expected for uniform GB energy. For heterogeneous GB energy distributions, the energy minimization principle dictates that high-energy interfaces should disappear faster than low-energy ones. The Het formulation, despite its widespread use, predicts the opposite trend: in both RS$^+$ and Bumpy cases, low-energy high-angle GBs are eliminated more rapidly than higher-energy interfaces (Fig.~\ref{Figure_4} (e)(f)), revealing a fundamental inconsistency. The inclusion of convective terms in HetGrad and HetGradProj partially restores the correct energetic bias. Contrary to the original modeling assumption, the ordinary gradient $\nabla\gamma$ in HetGrad (Fig.~\ref{Figure_4} (h)(i)) captures heterogeneity effects more effectively than the projected gradient employed in HetGradProj (Fig.~\ref{Figure_4} (k)(l)). The proposed formulation performs comparably to HetGrad in the case of RS$^+$ (Fig.~\ref{Figure_4} (n)). However, in strongly heterogeneous scenarios exemplified by the Bumpy case, it clearly outperforms all other models (Fig.~\ref{Figure_4} (o)). Not only the high-angle energy minimum near 60$^\circ$, but also the two bumps at $20^\circ$ and $40^\circ$, as well as the intermediate cusp, are accurately represented. Among the evaluated front-capturing level-set approaches, the proposed framework demonstrates the highest degree of energetic consistency in capturing heterogeneous GB kinetics.

\begin{figure}
    \centering
    \includegraphics[width=0.6\textwidth]{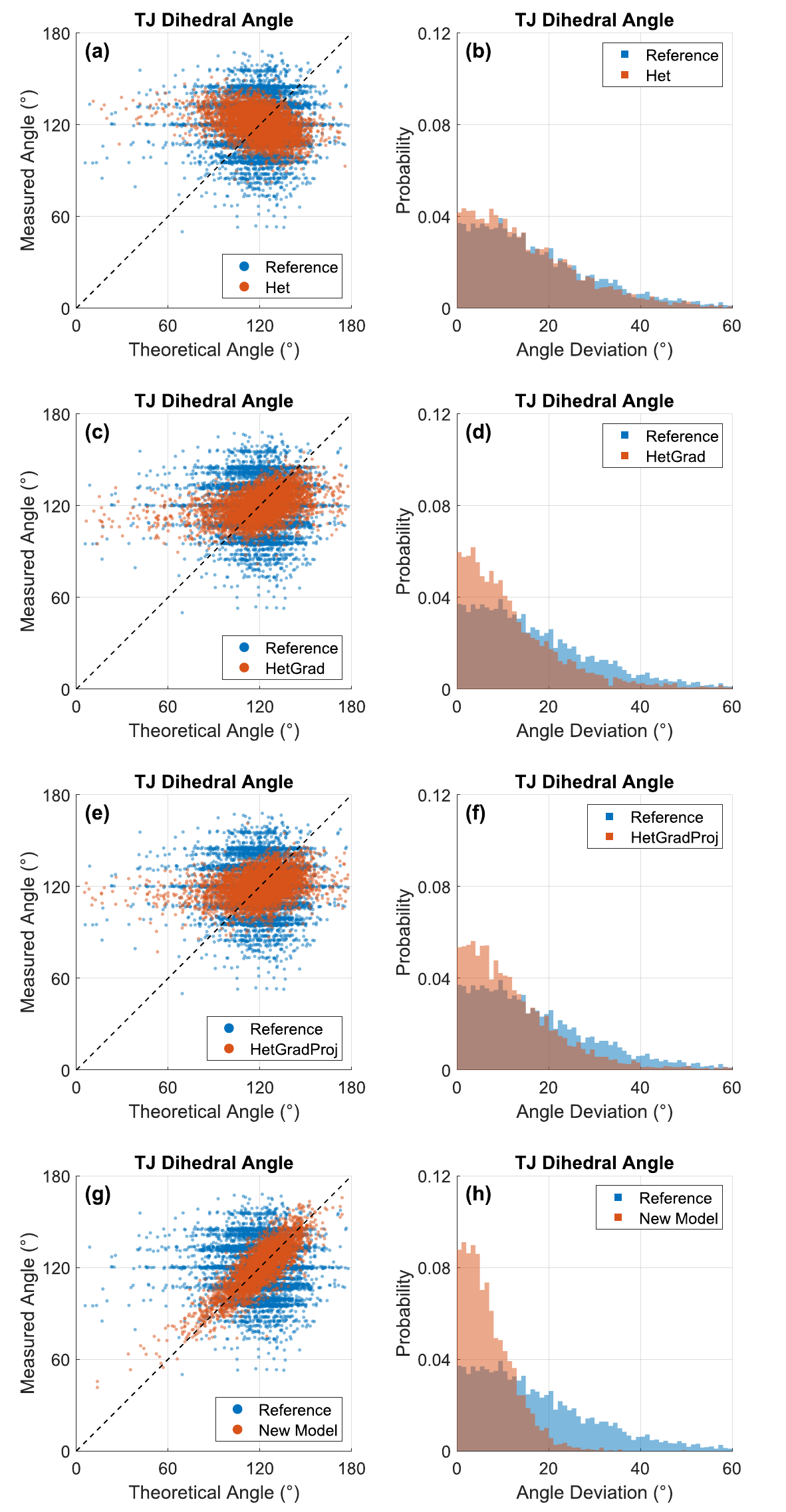}
    \caption{Statistics of triple junction dihedral angles in Bumpy simulations using \textbf{(a)}\textbf{(b)} Het, \textbf{(c)}\textbf{(d)} HetGrad, \textbf{(e)}\textbf{(f)} HetGradProj, and \textbf{(g)}\textbf{(h)} new model. \textbf{left column:} Measured angles plotted against their theoretical equilibrium values computed using Young's equation. \textbf{right column:} Deviation of measured angles from theoretical predictions.}
    \label{Figure_5}
\end{figure}

The superior performance of the proposed formulation stems from its accurate description of TJ kinetics during grain growth. As demonstrated in \cite{li_2025}, the capillarity-driven motion of an isolated TJ is reproduced with high fidelity over a wide range of GB energy ratios. In the present work, an analogous TJ analysis is conducted at the polycrystalline scale for all four level-set formulations. For the highly heterogeneous Bumpy case, dihedral angles of all TJs are measured at $T = 2$~h and compared to their theoretical equilibrium values computed from the local GB energies using Young’s equation (Fig.~\ref{Figure_5} (a)(c)(e)(g)). The deviations between measured and theoretical angles are further quantified (Fig.~\ref{Figure_5} (b)(d)(f)(h)). In the scatter plots, the dashed line denotes perfect agreement with theory, corresponding to $0^\circ$ deviation in the histograms. The proximity of data points to this line provides a direct measure of model accuracy. Among the existing approaches, the Het formulation exhibits the largest discrepancy: measured dihedral angles show little correlation with theoretical predictions (Fig.~\ref{Figure_5} (a)(b)). The HetGrad and HetGradProj models perform similarly. While they improve predictions under mild heterogeneity, their accuracy deteriorates significantly in strongly heterogeneous regimes, where theoretical equilibrium angles deviate markedly from $120^\circ$ (Fig.~\ref{Figure_5} (c)(d)(e)(f)). In contrast, the proposed formulation maintains accurate agreement with analytical dihedral angles across the entire heterogeneity spectrum, including extreme configurations approaching $0^\circ$ and $180^\circ$ (Fig.~\ref{Figure_5} (g)(h)). Nevertheless, it is worth noting that TJs have finite lifetimes in polycrystalline simulations. The junction interactions during grain growth lead to GB annihilation and reformation, altering local energy configurations and equilibrium angles. Many TJs are eliminated before reaching equilibrium, especially in fully-recrystallized fine-grained microstructures with high junction density. Furthermore, compared to the single-TJ benchmark in \cite{li_2025}, large-scale simulations necessarily employ coarser mesh near TJs to maintain computational efficiency, which slightly reduces dihedral angle measurement precision. These factors explain the remaining deviations from analytical predictions.

The proposed level-set formulation preserves its accuracy and robustness at the polycrystalline scale. Owing to its precise representation of TJ dihedral angles throughout evolving GB networks, it delivers the most energetically consistent grain growth kinetics among the level-set approaches examined in this work. Among the existing models, the widely used Het formulation exhibits the lowest overall accuracy. Incorporation of an energy gradient term in HetGrad improves predictions by partially accounting for local variations of GB energy at TJs. However, the gradient projection strategy employed in HetGradProj does not enhance predictive capability in the present polycrystalline simulations. While HetGradProj may yield improved TJ velocity in certain configurations, it does not provide superior dihedral angle accuracy. Since TJ dihedral angles are directly governed by the local energy balance through Young’s equation, their accurate prediction is essential for energetically consistent grain growth modeling. The proposed formulation prevails by capturing both TJ velocities and equilibrium dihedral angles with high fidelity.

The present framework demonstrates strong capability for modeling grain growth in highly heterogeneous polycrystals. It provides an efficient and physically consistent solution to long-standing challenges in computational metallurgy, enabling the annealing study of complex polycrystalline materials. Beyond its direct application, the high-fidelity level-set simulations produced by this framework may benefit other numerical approaches, including front-tracking methods \cite{mora_2008,florez_2020} and emerging deep-learning-based models \cite{bernacki_2025,tep_2025}. In particular, they can serve as reference, e.g., as ground-truth training datasets for data-driven grain growth models or as benchmarks for cross-validation with front-tracking simulations. 

Future developments aim to extend the current framework from 2D to 3D, by taking into account the quadruple junctions. A further objective is to incorporate anisotropic GB properties, such as inclination-dependent interface energy, through the introduction of GB stiffness and curvature tensors. The disconnection-based GB migration theory may also be considered to introduce physical insights based on microscopic mechanisms \cite{han_2018}. 

\section*{CRediT authorship contribution statement}

\textbf{Tianchi Li}: Software, Conceptualization, Methodology, Investigation, Formal analysis, Validation, Visualization, Writing – original draft, Writing – review \& editing

\textbf{Marc Bernacki}: Supervision, Software, Conceptualization, Writing – review \& editing, Funding acquisition, Project administration

\section*{Declaration of competing interest}

The authors declare that they have no known competing financial
interests or personal relationships that could have appeared to influence
the work reported in this paper.

\section*{Acknowledgements}

The authors thank Carnot M.I.N.E.S and ANR for their financial
support (Grant No. 230000488).

\bibliography{bibliography}

@article{humphreys_2017,
title={{Recrystallization and Related Annealing Phenomena}},
author={Humphreys, John and Rohrer, Gregory S. and Rollett, Anthony},
isbn={9780080982359},
edition = {3rd},
year={2017},
doi={},
publisher={Elsevier}
}

@article{smith_1948_rohrer,
author = {Rohrer, Gregory},
year = {2010},
month = {05},
pages = {1063-1100},
title = {{“Introduction to Grains, Phases, and Interfaces—an Interpretation of Microstructure”, Trans. AIME, 1948, vol. 175, pp. 15–51, by C.S. Smith}},
volume = {41},
journal = {Metallurgical and Materials Transactions A},
doi = {10.1007/s11661-010-0215-5}
}

@article{von_neumann_1952,
  title={Metal interfaces},
  author={Von Neumann, John},
  journal={American Society for Metals, Cleveland},
  volume={108},
  pages={108-110},
  year={1952}
}

@article{burke_1952,
title = {Recrystallization and grain growth},
journal = {Progress in Metal Physics},
volume = {3},
pages = {220-292},
year = {1952},
issn = {0502-8205},
doi = {10.1016/0502-8205(52)90009-9},
author = {J.E. Burke and D. Turnbull}
}

@article{mullins_1956,
author = {Mullins, W. W.},
title = {{Two‐Dimensional Motion of Idealized Grain Boundaries}},
journal = {Journal of Applied Physics},
volume = {27},
number = {8},
pages = {900-904},
year = {1956},
month = {08},
issn = {0021-8979},
doi = {10.1063/1.1722511}
}

@article{hillert_1965,
title = {On the theory of normal and abnormal grain growth},
journal = {Acta Metallurgica},
volume = {13},
number = {3},
pages = {227-238},
year = {1965},
issn = {0001-6160},
doi = {10.1016/0001-6160(65)90200-2},
author = {M. Hillert}
}

@article{lasalmonie_1986,
author={Lasalmonie, A. and Strudel, J. L.},
title={Influence of grain size on the mechanical behaviour of some high strength materials},
journal={Journal of Materials Science},
year={1986},
month={Jun},
day={01},
volume={21},
number={6},
pages={1837-1852},
issn={1573-4803},
doi={10.1007/BF00547918}
}

@article{zhang_2022,
title = {Study on creep properties of nickel-based superalloy blades based on microstructure characteristics},
journal = {Journal of Alloys and Compounds},
volume = {890},
pages = {161710},
year = {2022},
issn = {0925-8388},
doi = {10.1016/j.jallcom.2021.161710},
author = {Chengjiang Zhang and Ping Wang and Zhixun Wen and Zhuangzhuang Xu and Pengfei He and Zhufeng Yue}
}

@article{bhattacharya_2021,
author = {Aditi Bhattacharya  and Yu-Feng Shen  and Christopher M. Hefferan  and Shiu Fai Li  and Jonathan Lind  and Robert M. Suter  and Carl E. Krill  and Gregory S. Rohrer },
title = {Grain boundary velocity and curvature are not correlated in {Ni} polycrystals},
journal = {Science},
volume = {374},
number = {6564},
pages = {189-193},
year = {2021},
doi = {10.1126/science.abj3210}
}

@article{florez_2022,
title = {Statistical behaviour of interfaces subjected to curvature flow and torque effects applied to microstructural evolutions},
journal = {Acta Materialia},
volume = {222},
pages = {117459},
year = {2022},
issn = {1359-6454},
doi = {10.1016/j.actamat.2021.117459},
author = {Sebastian Florez and Karen Alvarado and Brayan Murgas and Nathalie Bozzolo and Dominique Chatain and Carl E. Krill and Mingyan Wang and Gregory S. Rohrer and Marc Bernacki}
}

@article{muralikrishnan_2023,
title = {Observations of unexpected grain boundary migration in {SrTiO3}},
journal = {Scripta Materialia},
volume = {222},
pages = {115055},
year = {2023},
issn = {1359-6462},
doi = {10.1016/j.scriptamat.2022.115055},
author = {Vivekanand Muralikrishnan and He Liu and Lin Yang and Bryan Conry and Christopher J. Marvel and Martin P. Harmer and Gregory S. Rohrer and Michael R. Tonks and Robert M. Suter and Carl E. Krill and Amanda R. Krause}
}

@article{xu_2024,
title = {Grain boundary migration in polycrystalline {$\alpha$}-{Fe}},
journal = {Acta Materialia},
volume = {264},
pages = {119541},
year = {2024},
issn = {1359-6454},
doi = {10.1016/j.actamat.2023.119541},
author = {Zipeng Xu and Yu-Feng Shen and S. Kiana Naghibzadeh and Xiaoyao Peng and Vivekanand Muralikrishnan and S. Maddali and D. Menasche and Amanda R. Krause and Kaushik Dayal and Robert M. Suter and Gregory S. Rohrer}
}

@article{yang_2026,
title = {Beyond curvature-driven grain growth: Insights from fully anisotropic {Monte Carlo Potts} simulations},
journal = {Acta Materialia},
volume = {302},
pages = {121672},
year = {2026},
issn = {1359-6454},
doi = {10.1016/j.actamat.2025.121672},
author = {Lin Yang and Vivekanand Muralikrishnan and Vishal Yadav and Zhihui Tian and Joel B. Harley and Amanda R. Krause and Michael R. Tonks}
}

@misc{bernacki_2025,
      title={{Predicting Grain Growth in Polycrystalline Materials Using Deep Learning Time Series Models}}, 
      author={Eliane Younes and Elie Hachem and Marc Bernacki},
      year={2025},
      eprint={2511.11630},
      archivePrefix={arXiv},
      primaryClass={cs.LG}
}

@Inbook{herring_1999,
author={Herring, Conyers},
title= {Surface Tension as a Motivation for Sintering},
bookTitle= {Fundamental Contributions to the Continuum Theory of Evolving Phase Interfaces in Solids: A Collection of Reprints of 14 Seminal Papers},
year ={1999},
publisher= {Springer Berlin Heidelberg},
address= {Berlin, Heidelberg},
pages= {33-69},
doi= {10.1007/978-3-642-59938-5_2}
}

@article{bernacki_2024,
title = {Kinetic equations and level-set approach for simulating solid-state microstructure evolutions at the mesoscopic scale: State of the art, limitations, and prospects},
journal = {Progress in Materials Science},
volume = {142},
pages = {101224},
year = {2024},
issn = {0079-6425},
doi = {10.1016/j.pmatsci.2023.101224},
author = {M. Bernacki}
}

@article{bulatov_2014,
title = {Grain boundary energy function for fcc metals},
journal = {Acta Materialia},
volume = {65},
pages = {161-175},
year = {2014},
issn = {1359-6454},
doi = {10.1016/j.actamat.2013.10.057},
author = {Vasily V. Bulatov and Bryan W. Reed and Mukul Kumar}
}

@Article{murgas_2021,
AUTHOR = {Murgas, Brayan and Florez, Sebastian and Bozzolo, Nathalie and Fausty, Julien and Bernacki, Marc},
TITLE = {{Comparative Study and Limits of Different Level-Set Formulations for the Modeling of Anisotropic Grain Growth}},
JOURNAL = {Materials},
VOLUME = {14},
YEAR = {2021},
NUMBER = {14},
ARTICLE-NUMBER = {3883},
PubMedID = {34300801},
ISSN = {1996-1944},
DOI = {10.3390/ma14143883}
}

@article{fausty_2018,
title = {A novel level-set finite element formulation for grain growth with heterogeneous grain boundary energies},
journal = {Materials \& Design},
volume = {160},
pages = {578-590},
year = {2018},
issn = {0264-1275},
doi = {10.1016/j.matdes.2018.09.050},
author = {Julien Fausty and Nathalie Bozzolo and Daniel {Pino Muñoz} and Marc Bernacki}
}

@article{fausty_2020,
title = {A 2{D} level set finite element grain coarsening study with heterogeneous grain boundary energies},
journal = {Applied Mathematical Modelling},
volume = {78},
pages = {505-518},
year = {2020},
issn = {0307-904X},
doi = {10.1016/j.apm.2019.10.008},
author = {Julien Fausty and Nathalie Bozzolo and Marc Bernacki}
}

@article{hallberg_2019,
doi = {10.1088/1361-651X/ab0c6c},
year = {2019},
month = {apr},
publisher = {IOP Publishing},
volume = {27},
number = {4},
pages = {045002},
author = {Hallberg, Håkan and Bulatov, Vasily V},
title = {Modeling of grain growth under fully anisotropic grain boundary energy},
journal = {Modelling and Simulation in Materials Science and Engineering}
}

@article{fausty_2021,
title = {A new analytical test case for anisotropic grain growth problems},
journal = {Applied Mathematical Modelling},
volume = {93},
pages = {28-52},
year = {2021},
issn = {0307-904X},
doi = {10.1016/j.apm.2020.11.035},
author = {J. Fausty and B. Murgas and S. Florez and N. Bozzolo and M. Bernacki}
}

@article{jin_2015,
title = {2{D} finite element modeling of misorientation dependent anisotropic grain growth in polycrystalline materials: Level set versus multi-phase-field method},
journal = {Computational Materials Science},
volume = {104},
pages = {108-123},
year = {2015},
issn = {0927-0256},
doi = {10.1016/j.commatsci.2015.03.012},
author = {Y. Jin and N. Bozzolo and A.D. Rollett and M. Bernacki}
}

@article{li_2025,
title = {An accurate and robust level-set formulation for multiple junction kinetics},
journal = {Scripta Materialia},
volume = {269},
pages = {116904},
year = {2025},
issn = {1359-6462},
doi = {10.1016/j.scriptamat.2025.116904},
author = {Tianchi Li and Marc Bernacki}
}

@article{scholtes_2015,
title = {New finite element developments for the full field modeling of microstructural evolutions using the level-set method},
journal = {Computational Materials Science},
volume = {109},
pages = {388-398},
year = {2015},
issn = {0927-0256},
doi = {10.1016/j.commatsci.2015.07.042},
author = {Benjamin Scholtes and Modesar Shakoor and Amico Settefrati and Pierre-Olivier Bouchard and Nathalie Bozzolo and Marc Bernacki}
}

@article{zhao_1996,
title = {{A Variational Level Set Approach to Multiphase Motion}},
journal = {Journal of Computational Physics},
volume = {127},
number = {1},
pages = {179-195},
year = {1996},
issn = {0021-9991},
doi = {10.1006/jcph.1996.0167},
author = {Hong-Kai Zhao and T. Chan and B. Merriman and S. Osher}
}

@article{hitti_2012,
title = {Precise generation of complex statistical {Representative Volume Elements (RVEs)} in a finite element context},
journal = {Computational Materials Science},
volume = {61},
pages = {224-238},
year = {2012},
issn = {0927-0256},
doi = {10.1016/j.commatsci.2012.04.011},
author = {K. Hitti and P. Laure and T. Coupez and L. Silva and M. Bernacki}
}

@article{mackenzie_1958,
    author = {Mackenzie, J. K.},
    title = {{Second Paper on Statistics Associated with the Random Disorientation of Cubes}},
    journal = {Biometrika},
    volume = {45},
    number = {1-2},
    pages = {229-240},
    year = {1958},
    month = {06},
    issn = {0006-3444},
    doi = {10.1093/biomet/45.1-2.229}
}

@article{read_shockley_1950,
  title = {{Dislocation Models of Crystal Grain Boundaries}},
  author = {Read, W. T. and Shockley, W.},
  journal = {Phys. Rev.},
  volume = {78},
  issue = {3},
  pages = {275--289},
  numpages = {0},
  year = {1950},
  month = {May},
  publisher = {American Physical Society},
  doi = {10.1103/PhysRev.78.275}
}

@article{digonnet_2007,
author = {Digonnet, Hugues and Silva, Luisa and Coupez, Thierry},
title = {{Cimlib: A Fully Parallel Application For Numerical Simulations Based On Components Assembly}},
journal = {AIP Conference Proceedings},
volume = {908},
number = {1},
pages = {269-274},
year = {2007},
month = {05},
doi = {10.1063/1.2740823}
}

@article{mesri_2009,
author = {Youssef Mesri, Hugues Digonnet and Thierry Coupez},
title = {Advanced parallel computing in material forming with {CIMLib}},
journal = {European Journal of Computational Mechanics},
volume = {18},
number = {7-8},
pages = {669--694},
year = {2009},
doi = {10.3166/ejcm.18.669-694}
}

@article{bernacki_2011,
title = {Level set framework for the finite-element modelling of recrystallization and grain growth in polycrystalline materials},
journal = {Scripta Materialia},
volume = {64},
number = {6},
pages = {525-528},
year = {2011},
issn = {1359-6462},
doi = {10.1016/j.scriptamat.2010.11.032},
author={Bernacki, Marc and Log{\'e}, Roland E and Coupez, Thierry}
}

@article{mora_2008,
title = {Three-dimensional grain growth: Analytical approaches and computer simulations},
journal = {Acta Materialia},
volume = {56},
number = {20},
pages = {5915-5926},
year = {2008},
issn = {1359-6454},
doi = {10.1016/j.actamat.2008.08.006},
author = {L.A. {Barrales Mora} and G. Gottstein and L.S. Shvindlerman}
}

@article{florez_2020,
title = {{A novel highly efficient Lagrangian model for massively multidomain simulation applied to microstructural evolutions}},
journal = {Computer Methods in Applied Mechanics and Engineering},
volume = {367},
pages = {113107},
year = {2020},
issn = {0045-7825},
doi = {10.1016/j.cma.2020.113107},
author = {Sebastian Florez and Karen Alvarado and Daniel Pino Muñoz and Marc Bernacki}
}

@article{tep_2025,
title = {High-fidelity grain growth modeling: Leveraging deep learning for fast computations},
journal = {Acta Materialia},
volume = {301},
pages = {121486},
year = {2025},
issn = {1359-6454},
doi = {10.1016/j.actamat.2025.121486},
author = {Pungponhavoan Tep and Marc Bernacki}
}

@article{han_2018,
title = {Grain-boundary kinetics: A unified approach},
journal = {Progress in Materials Science},
volume = {98},
pages = {386-476},
year = {2018},
issn = {0079-6425},
doi = {10.1016/j.pmatsci.2018.05.004},
author = {Jian Han and Spencer L. Thomas and David J. Srolovitz}
}

\end{document}